# Stereo image Transference & Retrieval over SMS

Muhammad Fahad Khan, Saira Beg

**Abstract**— Paper presents the way of transferring stereo images using SMS over GSM network. Generally, Stereo image is composed of two stereoscopic images in such way that gives three dimensional affect when viewed. GSM have two short messaging services, which can transfer images and sounds etc. Such services are known as; MMS (Multimedia Messaging Service) and EMS (Extended Messaging Service). EMS can send Predefined sounds, animation and images but have limitation that it does not support widely. MMS can send much higher contents than EMS but need 3G and other network capability in order to send large size data up to 1000 bytes. Other limitations are Portability, content adaption etc. Our major aim in this paper is to provide an alternative way of sending stereo images over SMS which is widely supported than EMS. We develop an application using J2ME Platform.

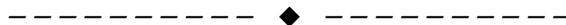

## 1 INTRODUCTION

In recent century, 3D technology has evolved. Different tools and methods are presents for displaying, retrieval, construction of stereo images and many more [1-2]. In GSM only EMS and MMS can transfer images and sounds. EMS is an application level extension of SMS, which can send predefined sounds, images and animations. All contents of EMS are present in message header, which will ignore in unsupported mobile phones [3-4].MMS can deliver richer contents such as; video, sounds, pictures, animation etc. It has lager data size up to 1000 bytes, delivering such a large size its need higher technology support such as 3G etc. some problem related to MMS are device compatibility, interoperability( not guaranteed), required upgraded messaging infrastructure, new billing structure, content adaption etc. [5].

This paper presents an alternative way of sending stereo images using SMS over GSM network. We developed an application which enables such features for all GSM devices (support SMS), even those which does not have EMS/MMS/GPRS/EDGE or other 3,4G capability. Section 2 is about related work, section 3, 4 presents proposed methodology and results respectively, and in last we conclude our paper.

## 2 RELATED WORK

Li et al [6] is about image displaying with cordless phone. An image display which can be capable of used with cordless phone such image display is a standalone image display device and have telephone and internet (wired or wireless connection). [7] discuss the making of stereoscopic images through mobile devices. In this they introduce the concept of single mobile device with multiple cameras. Each camera is capable of converging on a one point and captures standard images and that images will combine in order to create stereoscopic images. [12] presents the method of sending voice using SMS over GSM network. They used three formats for their results; PCM, ULAW and AMR. AMR format generates smaller number of messages, characters and connected messages than other two. Because it has audio data compression scheme in it.

## 3 PROPOSED METHODOLOGY

If sender wants to send a stereo image to receiver, he/she just browses and selects the required image. Our application running in a mobile phone takes that selected content (image) as input and applies different procedures in order to send it through SMS. First, it saves the selected content into signed ByteArrayOutputStream. After that, it converts signed ByteArrayOutputStream into unsigned integer array. The next step is to convert values of unsigned integer array into their respected Extended ASCII characters. But before such conversion adds 256 to all those values in array which falls in the range of 0-31 in order to move them on to the range 256 to 287 respectively. Main reason behind such addition is that, 0-31of ASCII characters are generally used for special functions such as; null etc. And such values cannot send through SMS. Lastly, it converts ASCII characters into strings and set those strings as a payload text of SMS as shown in figure1.

---

- *M. Fahad.Khan is with the Federal Urdu University of Arts, Science and Technology, Islamabad. mfahad.bs@gmail.com.*
- *Saira.Beg is with the COMSATS Institute of Information and Technology, Islamabad. sairabegbs@gmail.com*

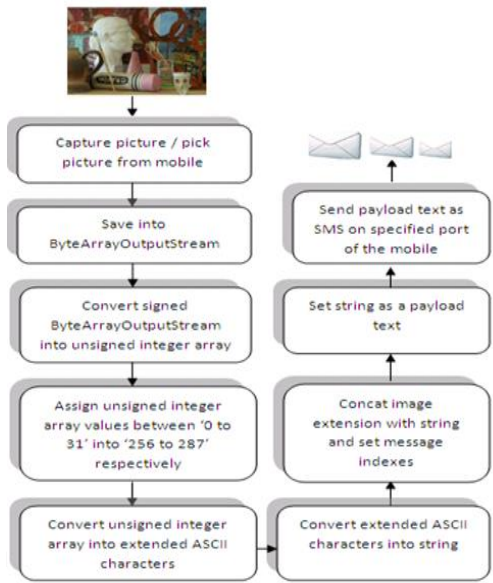

Fig. 1: Flow diagram of proposed methodology

In GSM, SMS have very small size up to 140 bytes, so concatenated SMS is used [8]. For sequence indexing (arrange the sequence of SMS) first 3 characters are reserved, such gives 000-999 connected SMS Indexing. When SMS is received at receiver side, first it place SMS in sequence with the help of index number and extracts its payload text and save it in RMS (Record Management System) as a single record. RMS can store record in bytes, so converts characters in bytes before storing [9]. And when receiver wants to see the image, then reverse procedure as described earlier is applied in real time. Figure 2 shows the interaction of proposed methodology within the existing GSM-SMS architecture. No change in existing architectur is required. We use onle SMS service of GSM.

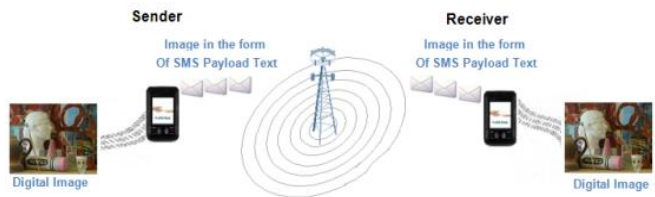

Fig. 2 represents the interaction within the existing GSM-SMS architecture

## 4 RESULTS AND DISCUSSION

We develop an application using J2ME platform and used Nokia N79 which supports JSR 184. And JSR 184 supports 3D images on mobile phones. During experiment we consider three main factors; Number of characters, number of messages and number of unique colors. We evaluate our application with two different tests. First we selected the six different stereo images from the [2, 10, 11] and calculate their results. Table 1 represents the six different images and table 2 shows the results respectively. Each image has same size and resolution but has different objects.

TABLE 1: REPRESENT STEREO IMAGES

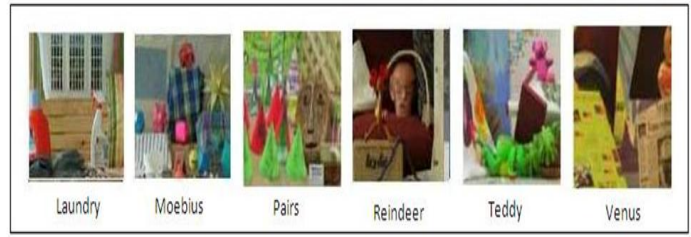

TABLE 2: RESULTS ACCORDING TO EACH IMAGE

| Data set | No. of Characters | No. of Messages | No. of unique colors |
|---|---|---|---|
| Laundry | 2244 | 6 | 2140 |
| Moebius | 2165 | 5 | 7940 |
| Pairs | 2140 | 5 | 8990 |
| Reindeer | 2115 | 5 | 7169 |
| Teddy | 2060 | 5 | 9017 |
| Venus | 1940 | 5 | 8159 |

In second phase, initially, we compare the affect of light in terms of messages, characters and colors. Table 3 shows the single image with six different light affects. Each image has same size and resolution but different light effects. Results of second phase is represented in figure 3 a, b and c.

TABLE 3: DIFFERENT LIGHT EFFECTS

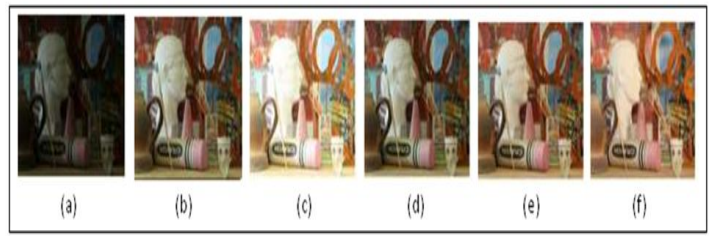

According to results we found that image "a" has least effect of light and "c" have maximum. Image "c" has higher number of messages then other images where "f" has higher number of characters and image "b" has higher number of unique colors.

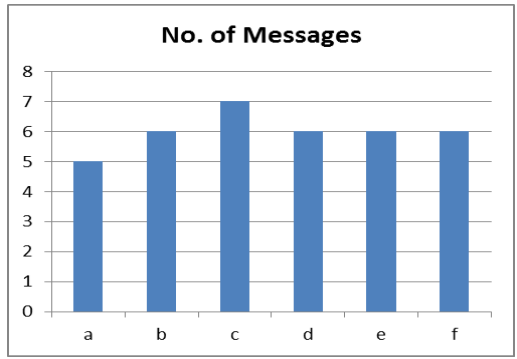

Fig. 3a: comparision of stereo images in terms of messages

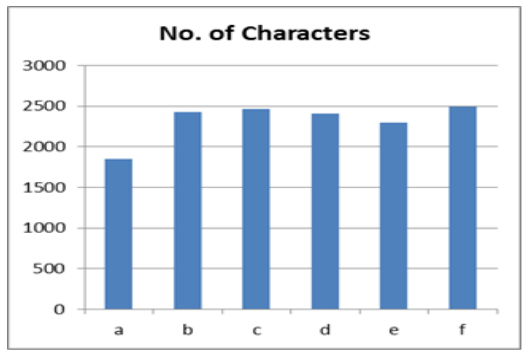

Fig. 3b: comparision of stereo images in terms of characters

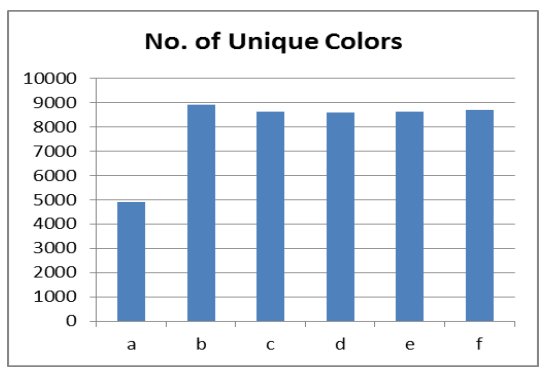

Fig. 3c: comparision of stereo images in terms of unique colors

## 5 CONCLUSION

Our major aim in this paper is to present an alternative way of sending stereo images using SMS over GSM network without capability of 3G and other higher networks and reduce hardware dependency. GSM have two short messaging services; EMS and MMS both can send images, sounds etc. but each have its own limitations where SMS cannot send images and sound and have limited size but still it is widely supported in all over the world. The biggest advantages of our application are its simplicity, and its ability to send images over GSM network without capability of higher data rate networks and utilize existing GSM-SMS architecture (no extra hardware is required).

Due to smaller size of SMS numbers of messages are higher which can be reduced by using compression algorithms.

## REFERENCES

[1] A. Ravishankar Rao and Alejandro Jaimes, "Digital Stereoscopic Imaging", Stereoscopic Displays and Applications X, IS&T/SPIE San Jose, CA Jan. 1999.
[2] Heiko Hirschmuller and Daniel Scharstein, " Evaluation of Cost Functions for Stereo Matching", In Proceedings of the IEEE Conference on Computer Vision and Pattern Recognition, June 2007
[3] Online available: http://www.developershome.com/sms/smsIntro.asp
[4] Online available: http://e-articles.info/e/a/title/The-Main-Protocols-used-by-Mobile-Phones-(SMS-EMS-MMS-WAP)/
[5] Daniel Ralph Paul Graham, "MMS Technologies, Usage, and Business Models", John Wiley AND Sons Ltd 2004, pp no 37, section 1.4.4, online available:
http://books.google.com.pk/books?id=PWv8cIcwC7YC&pg=PA4&lpg=PA4&dq=challenges+with+MMS&source=bl&ots=HV11xaZwNb&sig=FlOxT6Y3VIRtG54CXp0Qr0F4x28&hl=en&ei=Rd7zTaC3DYXLrQfvrdjhBg&sa=X&oi=book_result&ct=result&resnum=7&ved=0CFEQ6AEwBg#v=onepage&q=challenges&f=false
[6] LI et al. "image display with cordless phone", United States Patent Application, Publication Number US 2009/0128502 AI, May 21, 2009.
[7] Michael J. Chambers and Michael Kiessling, "Mobile communication device having stereoscopic image making capability" , United States Patent Application, Publication Number US2008/0064437 AI, Publication date 13 March 2008.
[8] Sun Microsystems, Inc "Wireless Messaging API (WMA) for Java™ 2 Micro Edition Reference Implementation", Version 1.0, JSR 120 Expert Groups, 2002.
[9] http://www.ibm.com/developerworks/library/wi-rms/
[10] http://vision.middlebury.edu/stereo/data/scenes2003/
[11] http://vision.middlebury.edu/stereo/data/scenes2005/
[12] M. Fahad Khan and Saira Beg, "Transferring Voice using SMS over GSM Network", Journal of Computing (JoC), Volume 3, Issue 4, PP 50-53, April 2011. ISSN-2151-9617

**MR. Muhammad Fahad Khan** is working as a Lecturer and In charge of Management Information System in Federal Urdu University of Arts, Science and Technology, Islamabad Pakistan. He has number of journal publications. He is reviewer of two international journals .He is interested in Handheld Application Development, Information System Development, Network Security, Digital Signal Processing and other their related fields. He did his Bachelor Degree from Federal Urdu University of Arts, Science and Technology in 2008 and now pursuing his Master Degree from IQRA University, Islamabad Campus

**MS. Saira Beg** is working as a Research Associate (May 2009- up to date) in COMSATS Institute of Information Technology, Islamabad. Her Interest areas are Networks, Network Security and Artificial Intelligence and other their related Fields. She is the member of Artificial Intelligence Group at CIIT, Islamabad. She did her Bachelor Degree (Gold medalist) from Federal Urdu University of Arts, Science and Technology, Islamabad, Pakistan in 2008 and now doing her Master Degree from COMSATS Institute of Information Technology, Islamabad.